\documentclass[usenatbib]{mnras}
\usepackage{txfonts, color,hyperref,graphicx,amssymb,latexsym,float}

\newcommand{\solarmass}{\,{M}_\odot}

\begin{document}

\title[The open flux evolution of a solar-mass star]{The open flux evolution of a solar-mass star on the main sequence}

\author[V. See et al.]
{V. See$^{1,2}$\thanks{E-mail: w.see@exeter.ac.uk}, M. Jardine$^{1}$, A. A. Vidotto$^{3}$, J.-F. Donati$^{4,5}$, S. Boro Saikia$^{6,7}$, R. Fares$^{8,9}$,
\newauthor C. P. Folsom$^{4,5,10,11}$, S. V. Jeffers$^{6}$, S. C. Marsden$^{12}$, J. Morin$^{13}$, P. Petit$^{4,5}$,
\newauthor and the BCool Collaboration\\
$^{1}$SUPA, School of Physics and Astronomy, University of St Andrews, North Haugh, KY16 9SS, St Andrews, UK\\
$^{2}$Department of Physics and Astronomy, University of Exeter, Physics Building, Stocker Road, Exeter, EX4 4QL, UK\\
$^{3}$School of Physics, Trinity College Dublin, University of Dublin, Dublin-2, Ireland\\
$^{4}$Universit\'{e} de Toulouse, UPS-OMP, Institut de Recherche en Astrophysique et Plan\'{e}tologie, Toulouse, France\\
$^{5}$CNRS, Institut de Recherche en Astrophysique et Plan\'{e}tologie, 14 Avenue Edouard Belin, F-31400 Toulouse, France\\
$^{6}$Universit\"at G\"ottingen, Institut f\"ur Astrophysik, Friedrich-Hund-Platz 1, 37077 G\"ottingen, Germany\\
$^{7}$Institut f\"ur Astronophysik, Universit\"at Wien, T\"urkenschanzstrasse 17, 1180 Wien, Austria\\
$^{8}$INAF- Osservatorio Astrofisico di Catania, Via Santa Sofia, 78 , 95123 Catania, Italy\\
$^{9}$ Department of Natural Sciences, School of Arts and Sciences, Lebanese American University, P.O. Box 36, Byblos, Lebanon\\
$^{10}$Univ. Grenoble Alpes, IPAG, F-38000 Grenoble, France\\
$^{11}$CNRS, IPAG, F-38000 Grenoble, France \\
$^{12}$University of Southern Queensland, Computational Engineering and Science Research Centre, Toowoomba, 4350, Australia\\
$^{13}$Laboratoire Univers et Particules de Montpellier, Universit\'e de Montpellier, CNRS, F-34095, France\\}

\maketitle

\begin{abstract}
Magnetic activity is known to be correlated to the rotation period for moderately active main sequence solar-like stars. In turn, the stellar rotation period evolves as a result of magnetised stellar winds that carry away angular momentum. Understanding the interplay between magnetic activity and stellar rotation is therefore a central task for stellar astrophysics. Angular momentum evolution models typically employ spin-down torques that are formulated in terms of the surface magnetic field strength. However, these formulations fail to account for the magnetic field geometry, unlike those that are expressed in terms of the open flux, i.e. the magnetic flux along which stellar winds flow.

In this work, we model the angular momentum evolution of main sequence solar-mass stars using a torque law formulated in terms of the open flux. This is done using a potential field source surface model in conjunction with the Zeeman-Doppler magnetograms of a sample of roughly solar-mass stars. We explore how the open flux of these stars varies with stellar rotation and choice of source surface radii. We also explore the effect of field geometry by using two methods of determining the open flux. The first method only accounts for the dipole component while the second accounts for the full set of spherical harmonics available in the Zeeman-Doppler magnetogram. We find only a small difference between the two methods, demonstrating that the open flux, and indeed the spin-down, of main sequence solar-mass stars is likely dominated by the dipolar component of the magnetic field.
\end{abstract}

\begin{keywords} techniques: polarimetric - stars: activity - stars: evolution - stars: magnetic field - stars: rotation
\end{keywords}

\section{Introduction}
\label{sec:Intro}
Understanding how the angular momentum and rotation periods of low-mass stars ($M_{\star}\lesssim 1.3\solarmass$) evolve over their lifetimes is an important goal within stellar astrophysics. For example, rotation is known to be correlated to numerous forms of magnetic activity including X-ray emission \citep{Pizzolato2003,Wright2011}, chromospheric activity \citep{Noyes1984,Mamajek2008} and large-scale magnetic field generation \citep{Petit2008,Vidotto2014Trends,See2015,Folsom2016,See2016}. The stellar rotation period can also be used as a proxy for the stellar age using the so called gyrochronology relations \citep{Barnes2003,Barnes2007,Barnes2010}, at least for stars whose rotation periods have converged onto a single track in the rotation period - age plane (for solar mass stars, convergence occurs at $\sim$1 Gyr). Finally, the magnetic activity and rotation history of a host star can also have a significant impact on the potential habitability of exoplanets \citep{Wood2014,Tu2015,Gallet2016,Ribas2016}. For example, stellar winds can significantly compress planetary magnetospheres \citep{Vidotto2013,See2014} and reduce their ability to protect the planetary atmosphere from the erosive effects of the wind. Planetary atmospheres can also be eroded away by photoevaporation caused by high energy radiation \citep{Lammer2003}. The rate at which this occurs depends strongly on the initial rotation period of the host star \citep{Johnstone2015Atmospheres}.

Along the main sequence, the main agent of angular momentum evolution is magnetised stellar winds that carry angular momentum away from the central star. Many authors have studied the rate at which stars lose angular momentum \citep{Vidotto2014Torque,Cohen2014,Garraffo2015,Nicholson2016,See2017} or formulated braking laws that describe how the angular momentum loss varies as a function of the parameters of the host star \citep{Reiners2012,Matt2012,Reville2015}. These braking laws have subsequently been used to model stellar rotation period evolution from the pre-main sequence to ages older than the Sun \citep{Gallet2013,Brown2014,Gallet2015,Matt2015,Johnstone2015,vanSaders2016,Amard2016,Blackman2016}. 

Numerous studies have shown that the open flux is an important parameter in the context of angular momentum loss \citep{Mestel1987,Vidotto2014Torque,Reville2015,Reville2015b}. However, it is not directly observable and must be estimated using physical models. One such model is the potential field source surface (PFSS) model \citep{Altschuler1969}. This model takes an input magnetogram of the stellar surface and extrapolates the field upwards to the so called source surface; a spherical surface that represents the limit of coronal confinement. Once the source surface is reached, the field lines are assumed to be open and radial, mimicking the action of plasma pressure opening up closed field lines.

A number of factors can affect the amount of open flux estimated by the PFSS model. The first is the choice of input magnetogram. Previous theoretical work has shown that, when considering individual field modes, stars with dipolar surface fields have the most open flux and that the open flux decreases with increasing spherical harmonic degree (quadrupole, octupole, etc.) \citep{Reville2015,Garraffo2015}. For stellar studies, the input magnetogram is typically a Zeeman-Doppler imaging (ZDI) map \citep{Jardine2002,Gregory2006,Fares2010,Lang2012,Johnstone2014,See2015Radio}. ZDI is a tomographic technique that is capable of reconstructing the large-scale surface magnetic field structure of cool-dwarf stars \citep{Semel1989,Brown1991,Donati1997,Donati2006}. Over the last two decades, a considerable amount of effort has been dedicated to investigating the field geometry of low-mass stars. It has been found that their surface fields are composed of a mixture of spherical harmonic modes \citep[e.g.][]{Jeffers2014,Saikia2015,Saikia2016,Folsom2016}. However, recent work suggests that the open flux is dominated by the dipolar component of the field, at least for the choice of source surface radius used in those works \citep{Lang2014,Jardine2017,See2017}. This is because the dipolar component of the field decays the most slowly with height above the stellar surface. Given that the ZDI technique can typically reconstruct the stellar magnetic field up to a spherical harmonic mode of, at least, $\ell =5$, ZDI maps are an appropriate choice of inner boundary condition for the PFSS model in the context of determining open flux.

The source surface radius is another parameter that can affect the amount of open flux recovered. Within the PFSS model, it is a free parameter but it is observationally unconstrained for stars other than our Sun. For a given input map, more of the flux is forced to be open for smaller values of the source surface radius. Additionally, if the source surface is sufficiently small, the higher order field modes may not have completely decayed away and may contribute towards the open flux. Typically, the source surface radius is picked to have values similar to the solar value ($\sim 2.5 r_{\star}$) but in reality it should vary as a function of the fundamental parameters of the star \citep{Reville2015b}. 

In \citet{See2017}, we studied how the open flux and the corresponding spin-down torque varied using a sample of low-mass stars with a wide range of masses and rotation periods. We found that the open flux of stars with Rossby numbers, $\rm Ro$, greater than $\sim$0.01 follow the classical activity rotation relation shape but that $\rm Ro \lesssim 0.01$ stars departed from this relation. These results were obtained using the simplifying assumption that all the stars had source surface radii of $r_{\rm ss} = 3.41 r_{\star}$. This is a useful assumption since it allows for a rapid assessment of how stellar open flux varies over a large portion of the HR diagram. However, it ignores the fact that the source surface radius likely varies as a function of mass and rotation period. Indeed, to these authors' knowledge, there has not been a systematic study of how the source surface affects the open flux recovered for a set of realistic input magnetograms to date. 

In this study, we will use a sample of 22 main sequence stars of roughly solar-mass ($0.9\solarmass<M_{\star}<1.1\solarmass$) that have had their large-scale surface magnetic fields mapped to investigate the open flux evolution main sequence solar-mass stars. Using a PFSS model, we investigate how the open flux of these stars varies for different source surface radii and the effect of including/excluding higher order field modes. The angular momentum evolution of a solar-mass star can then be calculated over its lifetime using these open flux formulations, in conjunction with the braking law of \citet{Reville2015}. We use rotation period data from open clusters of known ages to constrain our angular momentum evolution model and determine how the source surface radius and open flux varies over the main sequence lifetime. In section \ref{sec:Model}, we outline the details of our spin-down model. In section \ref{sec:OpenFlux}, we outline two methods of determining the open flux as a function of rotation and source surface radius. In section \ref{sec:Clusters} we discuss the open cluster data we use to calibrate our model. In section \ref{sec:Tracks}, we present the results of our angular momentum evolution model. A discussion and the conclusions follow in section \ref{sec:Discussion}.

\section{Angular momentum evolution model}
\label{sec:Model}
In order determine how the rotation period of a star evolves, we need to solve the angular momentum equation, 

\begin{equation}
	\frac{d \Omega_{\star}}{dt} = \frac{\dot{J}}{I_{\star}} - \frac{\Omega_{\star}}{I_{\star}}\frac{dI_{\star}}{dt},
	\label{eq:OmegEvo}
\end{equation}
where $\Omega_{\star}=2\pi/P_{\rm rot}$ is the stellar angular velocity, $P_{\rm rot}$ is the stellar rotation period, $t$ is time, $\dot{J}$ is the angular momentum-loss rate or spin-down torque and $I_{\star}$ is the moment of inertia of the star. For simplicity, we will assume solid body rotation throughout the entire main sequence lifetime. We use the evolutionary models of \citet{Baraffe2015} for a solar-mass star to determine how the moment of inertia changes with time although we note that the internal structure of a star remains relatively constant on the main sequence and so the changes in angular velocity are dominated by the spin-down torque term in equation \ref{eq:OmegEvo}. 

\begin{figure}
	\begin{center}
	\includegraphics[trim=1cm 1cm 1cm 0cm,width=\columnwidth]{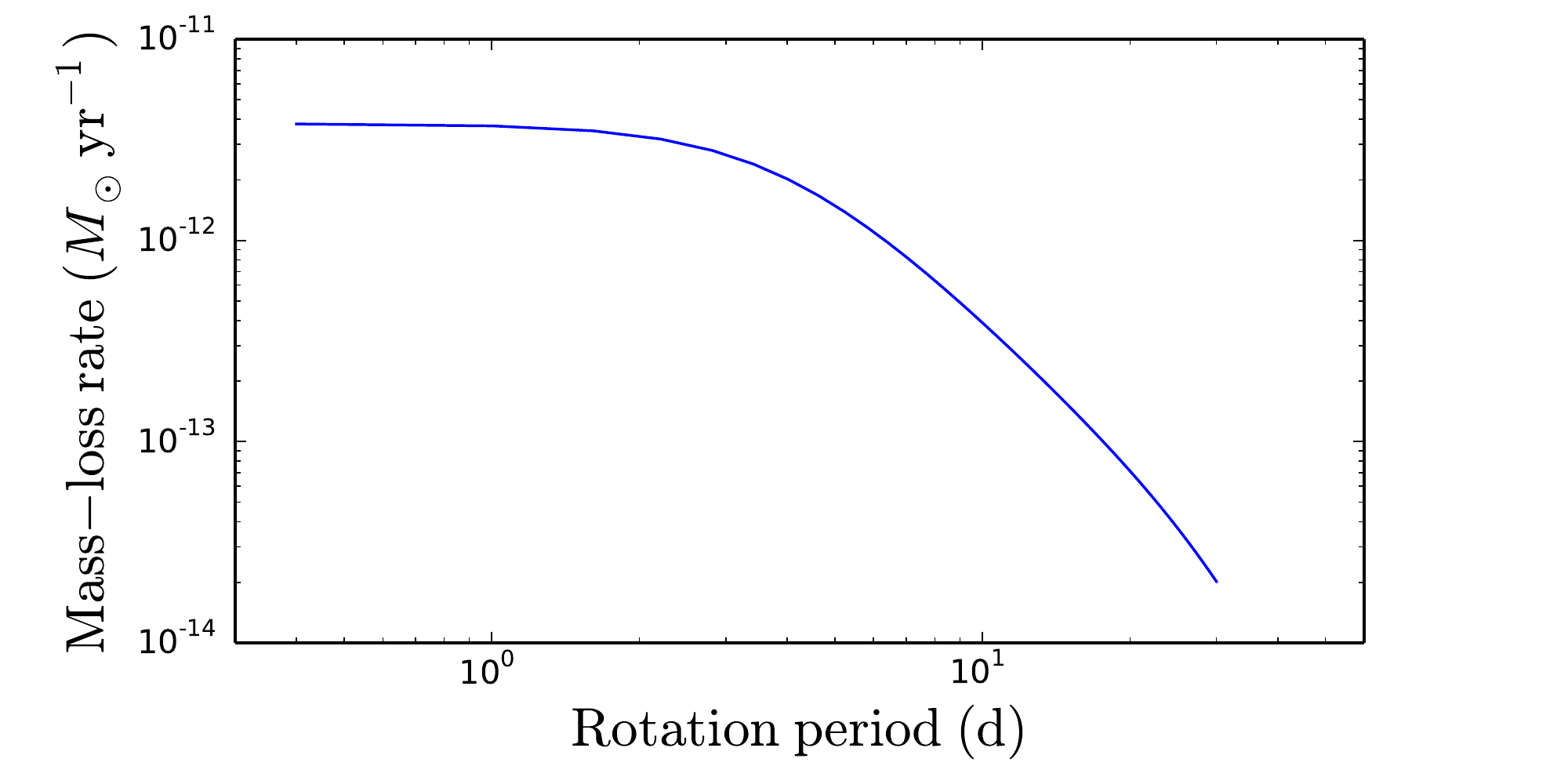}
	\end{center}
	\caption{Mass-loss rate against rotation period for a solar-mass star as calculated using the model of \citet{Cranmer2011}.}
	\label{fig:MassLoss}
\end{figure}

For the spin-down torque, we use the formulation of \citet{Reville2015}, 

\begin{equation}
	\dot{J}_{\rm R15}=\dot{M}\Omega_{\star}r_{\star}^2K_3^2\left(\frac{\Upsilon_{\rm open}}{\left(1+f^2/K_4^2\right)^{1/2}}\right)^{2m},
	\label{eq:jDot}
\end{equation}
where $\dot{M}$ is the mass-loss rate, $r_{\star}$ is the stellar radius, $\Upsilon_{\rm open}=\Phi^2_{\rm open}/(r_{\star}^2\dot{M}v_{\rm esc})$ is a measure of the magnetisation of the open field lines, $\Phi_{\rm open}$ is the open flux, $v_{\rm esc}=(2GM_{\star}/r_{\star})^{1/2}$ is the stellar escape velocity, $M_{\star}$ is the stellar mass, $f=\Omega_{\star}r_{\star}^{3/2}(GM_{\star})^{-1/2}$ is the angular rotation speed normalised to the breakup speed and $K_3=0.65$, $K_4=0.06$ and $m=0.31$ are fit parameters determined from the results of MHD simulations\footnote{The value of $K_3$ is given as 1.4 by \citet{Reville2015}. However, this is a typographical error and the true value is $K_3=0.65$ (R\'{e}ville, priv. comm.).}. We use the model of \citet{Cranmer2011} to estimate the mass-loss rate. This is a 1D model that estimates the magnitude of the alfv\'{e}n wave energy flux generated by subsurface convective motions. The model tracks this energy flux through the stellar surface and estimates the amount that is deposited as heat in the transition region. It is this heat that is responsible for driving the winds in this model. Many previous studies have shown that magnetic activity scales with rotation. In the model of \citet{Cranmer2011}, this scaling is encapsulated by the magnetic filling factor which is estimated using empirical scaling relations based on previously published data. Fig. \ref{fig:MassLoss} shows the mass-loss rate of a solar-mass star as a function of rotation period using this model. Similarly to other phenomena related to magnetic activity, the mass-loss rate increases with decreasing rotation period and saturates at the fastest rotation periods. In order to estimate the open flux, we will use the PFSS model in conjunction with ZDI maps. Investigating how the open flux varies with rotation and source surface radii will form the focus of this work and will be presented in section \ref{sec:OpenFlux}. Having calculated $\dot{J}_{\rm R15}$, we will assume that the real spin-down torque is proportional to this value, i.e. $\dot{J}=k\dot{J}_{\rm R15}$. Such an assumption has also been made in previous works \citep{Gallet2013,Gallet2015,Johnstone2015}. We will fit for the proportionality constant, $k$, and discuss its physical significance in section \ref{sec:Tracks}. Lastly, table \ref{tab:Values} contains the solar values that we use in this study.

\begin{table}
\begin{minipage}{70mm}
	\begin{center}	
	\caption{Adopted solar values in this study} 
	\label{tab:Values}
	\begin{tabular}{lc}
		\hline
		\hline
$M_{\odot}$ (solar mass)	&	$2.0 \times 10^{33}\ {\rm g}$	\\
$r_{\odot}$ (solar radius)	&	$6.96 \times 10^{10}\ {\rm cm}$	\\
$P_{\rm rot,\odot}$ (solar rotation period)	&	26 d	\\
$\tau_{\rm c}$ (solar-mass convective turnover time)	&	14.45 d	\\
$P_{\rm rot,crit}$ (critical rotation period)	&	1.45 d	\\
$r_{\rm ss,\odot}$ (solar source surface radius)	&	2.5 $r_{\odot}$	\\
$t_{\odot}$ (solar age)	&	4.6 Gyr	\\
	\hline
\end{tabular}
\end{center}
\end{minipage}
\end{table}

\section{Estimating the open flux}
\label{sec:OpenFlux}
\subsection{The potential field source surface model}
\label{subsec:PFSS}
In order to estimate the open flux, we use the PFSS model \citep{Altschuler1969}. In this model, the magnetic field is assumed to be in a potential state, i.e. current free, and the three components are given by 

\begin{equation}
	B_{\rm r}=-\sum\limits_{l=1}^N \sum\limits_{m=-l}^l [la_{lm} r^{l-1} - \left(l+1\right) b_{lm} r^{-\left(l+2\right)}] P_{lm} \left(\cos \theta \right) e^{im\phi}
	\label{eq:Br}
\end{equation}

\begin{equation}
	B_{\theta}=-\sum\limits_{l=1}^N \sum\limits_{m=-l}^l [a_{lm}r^{l-1} + b_{lm} r^{-\left(l+2\right)}]\frac{d}{d\theta} P_{lm} \left(\cos \theta \right) e^{im\phi}
	\label{eq:Btheta}
\end{equation}

\begin{equation}
	B_{\phi}=-\sum\limits_{l=1}^N \sum\limits_{m=-l}^l [a_{lm}r^{l-1} + b_{lm} r^{-\left(l+2\right)}] P_{lm} \left(\cos \theta\right) \frac{im}{\sin \theta} e^{im\phi}.
	\label{eq:Bphi}
\end{equation}
where $l$ is the spherical harmonic degree, $m$ is the azimuthal number, $a_{lm}$ \& $b_{lm}$ are the amplitudes of each spherical harmonic component and $P_{lm}$ are the Legendre polynomials. Equations (\ref{eq:Br}) - (\ref{eq:Bphi}) only apply between the stellar surface and the source surface with the field assumed to decay radially as an inverse square law above the source surface. In order to determine the values of $a_{lm}$ and $b_{lm}$, two boundary conditions are required. The first is the field geometry at the stellar surface which is set using ZDI maps. We use a sample of 22 main-sequence stars with masses between 0.9$\solarmass$ and 1.1$\solarmass$\footnote{There are a number of reported masses for AB Dor in the literature. \citet{Donati2003DR} give its mass as 1$\solarmass$ which falls within our mass bracket (0.9$\solarmass \leq 1\solarmass \leq 1.1\solarmass$). On the other hand, \citet{Guirado2010} report a dynamical mass of 0.86$\solarmass$. It is therefore unclear whether AB Dor should be included in our sample. We note that AB Dor falls into the saturated regime and is not included in any of the fits in this work. Its inclusion does not affect our final numerical results and it simply serves to show that the field flux values we obtain in the saturated regime are reasonable. We have therefore included it in this work.}. The large-scale surface magnetic fields of each of these stars have been mapped with ZDI, sometimes over multiple epochs. The parameters of these stars are shown in table \ref{tab:ZDIStars} along with a reference to the article their magnetic map was originally published in. The stellar parameters are taken from the same references or \citet{Vidotto2014Trends} and references therein. It should be noted that values for the stellar parameters have been obtained from a number of different sources which will add systematic consistency errors. The second boundary condition is the requirement that the field must become purely radial at some distance away from the star known as the source surface radius, $r_{\rm ss}$. As discussed in the introduction, the value of $r_{\rm ss}$ is a free parameter in this model. The open flux is then given by integrating the radial field over the source surface,

\begin{equation}
	\Phi_{\rm open} = \oint_{r_{\rm ss}} | B_r|\ {\rm d}S.
	\label{eq:Open}
\end{equation}

\citet{See2017} showed that for low-mass stars ($<1.4\solarmass$), the open flux is determined predominantly by the dipolar component of the magnetic field, at least for their chosen source surface radii of $r_{\rm ss} = 3.41R_{\star}$. In this work, we will investigate whether this assumption holds for different choices of the source surface radii. We will outline two methods of estimating the open flux of a solar mass star. The first method will use only the dipolar component of the ZDI maps as inputs to the PFSS model while the second method will use the full set of spherical harmonics available in ZDI maps. We will refer to these as the dipolar and the multipolar methods respectively.

\subsection{Dipolar method of determining open flux}
\label{subsec:DipOpenFlux}
\begin{figure}
	\begin{center}
	\includegraphics[trim=1cm 1cm 1cm 0cm,width=\columnwidth]{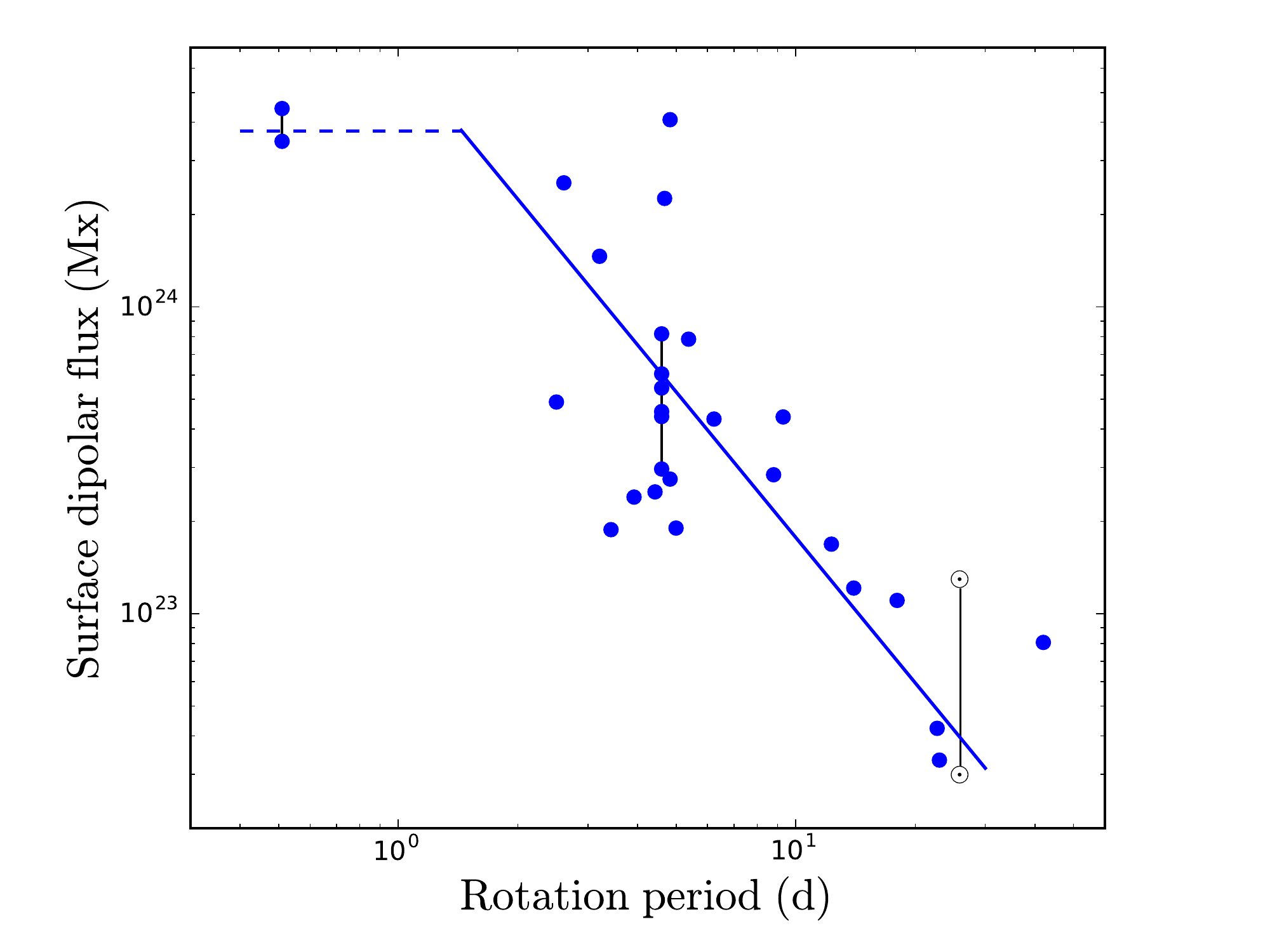}
	\end{center}
	\caption{Surface dipolar flux against rotation period. Stars observed at multiple epochs are connected by a vertical line. The fit to the stars in the unsaturated regime (solid blue line, $P_{\rm rot}>1.44\ {\rm d}$) has the form $\Phi_{\rm \star, dip}=(6.69\pm 3.28)\times 10^{24}P_{\rm rot}^{-1.58\pm 0.23}$. The surface dipolar flux has a value of $3.72\times 10^{24}\ {\rm Mx}$ in the saturated regime (dashed blue line, $P_{\rm rot}<1.44\ {\rm d}$). Two solar symbols indicate the typical variation of the surface dipolar flux over the solar cycle (values are taken from \citet{Jardine2017}).}
	\label{fig:SurDipFlux}
\end{figure}

Fig. \ref{fig:SurDipFlux} shows the surface flux associated with the dipolar component ($l=1$) of the ZDI maps, $\Phi_{\rm \star, dip}$, as a function of rotation period. In this work, we will define the unsaturated regime to be $\rm Ro>0.1$. For solar mass stars, which have convective turnover times of $14.45\ {\rm d}$ (calculated using equation (11) of \citealt{Wright2011}), this corresponds to a critical rotation period of $P_{\rm rot, crit} = 1.45\ {\rm d}$. The fit to the unsaturated stars (those stars with $P_{\rm rot}>P_{\rm rot, crit}$) has the form $\Phi_{\rm \star, dip}=(6.69\pm 3.28)\times 10^{24}P_{\rm rot}^{-1.58\pm 0.23}$. This fit, as well as the others in this work, was done using the bisector ordinary least-squares method \citep{Isobe1990}. The errors on the fit are determined by considering only the scatter of the points. From the presently available data, it is difficult to constrain the value  of the surface dipolar flux in the saturated regime. Currently, only a single star of roughly solar mass in the saturated regime (AB Dor) has been mapped with ZDI. For now we will assume that the surface dipolar flux transitions continuously from the unsaturated regime to the saturated regime such that it has a value of $6.69\times 10^{24}P_{\rm rot,crit}^{-1.58}=3.72\times10^{24}\ {\rm Mx}$. This corresponds well with the surface dipolar flux of AB Dor. However, further ZDI observations of solar-mass stars in the saturated regime will be required to determine if this estimate of dipolar flux in the saturated regime is representative of the true value.

In general, equations (\ref{eq:Br}) \& (\ref{eq:Open}) should be used to calculate the open flux when using all the available spherical harmonics in a ZDI map. However, when considering only the dipolar component of the ZDI map, the open to surface flux ratio is simply given by

\begin{equation}
	\frac{\Phi_{\rm open, dip}}{\Phi_{\rm\star ,dip}} = \frac{3\tilde{r}_{\rm ss}^2}{2\tilde{r}_{\rm ss}^3+1},
	\label{eq:OpenFluxRatio}
\end{equation}
where $\tilde{r}_{\rm ss}=r_{\rm ss}/r_{\star}$ is the ratio of the source surface radii to the stellar radii. This simple expression exists because we are dealing with only a single spherical harmonic mode and no cross terms arise when calculating the integral in equation (\ref{eq:Open}). A full derivation of equation (\ref{eq:OpenFluxRatio}) is available in appendix \ref{app:Dipole}. By combining equation (\ref{eq:OpenFluxRatio}) with the expression for the surface dipolar flux, as a function of rotation period, derived from Fig. \ref{fig:SurDipFlux}, we have a simple way of estimating the dipolar open flux of a star as a function of rotation period and source surface radii. In this work, we will assume that the source surface radius can be parameterised as 

\begin{equation}
	r_{\rm ss} = r_{\rm ss,\odot}\left(\frac{P_{\rm rot}}{P_{\rm rot,\odot}}\right)^n, \hspace{3mm} (P_{\rm rot}>P_{\rm rot, crit})
	\label{eq:rss}
\end{equation}
and
\begin{equation}
	r_{\rm ss} = r_{\rm ss,\odot}\left(\frac{P_{\rm rot, crit}}{P_{\rm rot,\odot}}\right)^n,  \hspace{3mm} (P_{\rm rot}<P_{\rm rot, crit}),
	\label{eq:rss2}
\end{equation}
where values for the solar source surface radius, $r_{\rm ss, \odot}$ and rotation period, $P_{\rm rot,\odot}$, can be found in table \ref{tab:Values} and $n$ is a power law index that we will determine in section \ref{sec:Tracks}. In principle, to properly determine the source surface radius, one should consider the location where the plasma thermal pressure and bulk ram pressure of the wind are able to overcome the magnetic pressure of the field causing the field lines to open up (see \citep{Reville2015b} for an in depth discussion of how to determine the source surface radius). However, we choose to use the simplified form presented in equations (\ref{eq:rss}) and (\ref{eq:rss2}). Such a dependence is not unreasonable given that many phenomena associated with stellar activity, such as large-scale magnetic fields \citep{Vidotto2014Trends,See2015}, mass-loss \citep{Cranmer2011} and X-ray emission \citep{Wright2011}, have a power law dependence on rotation in the unsaturated regime with saturation occurring for the fastest rotators. It should also be noted that the source surface radius of a given star should vary due to various forms of intrinsic variability in the stellar magnetic field such as stellar cycles. These short term fluctuations are not considered by our model and the value calculated using equations (\ref{eq:rss}) and (\ref{eq:rss2}) should be regarded as a source surface radius value averaged over evolutionary time-scales. 

\begin{table*}
\begin{minipage}{160mm}
	\begin{center}	
	\caption{Parameters of the stars mapped with ZDI listed in ascending order of rotation period: stellar mass, radius, rotation period, age, surface dipolar flux, total surface flux, the source surface radius for which the open flux calculated using the multipolar method differs from the open flux calculated using the dipolar method by 10\% (see sections \ref{subsec:DipOpenFlux} and \ref{subsec:FullOpenFlux}). No value is listed for stars for which the dipolar and multipolar open flux never differ by more than 10\%. The observation epoch at which each star was observed and references for the paper in which the original magnetic map was published are also listed. The stellar parameters are also taken from these references or \citet{Vidotto2014Trends} and references therein.} 
	\label{tab:ZDIStars}
	\begin{tabular}{lccccccccc}
	\hline
	Star & $M_{\star}$ & $r_{\star}$ & $P_{\rm rot}$ & Age & $\Phi_{\rm \star,dip}$ & $\Phi_{\star}$ & $r_{\rm ss, 10\%}$ & Obs & Ref.\\
	ID & ($M_{\odot}$) & ($r_{\odot}$) & (d) & [Myr] & ($10^{23}$ Mx) & ($10^{23}$ Mx) & ($r_{\star}$) & epoch & \\
	\hline
AB Dor	&	1$^a$	&	1	&	0.51	&	120	&	34.65	&	79.76	&	1.84	&	2001 Dec	&	\citet{Donati2003Dynamo}	\\
...	&	...	&	...	&	...	&	...	&	44.31	&	65.15	&	1.45	&	2002 Dec	&	...	\\
PELS 031	&	0.95	&	1.05	&	2.5	&	125	&	4.90	&	11.65	&	2.51	&	2013 Nov	&	\citet{Folsom2016}	\\
HII 296	&	0.9	&	0.93	&	2.61	&	125	&	25.72	&	27.69	&	-	&	2009 Oct	&	\citet{Folsom2016}	\\
BD-16351	&	0.9	&	0.88	&	3.21	&	42	&	14.62	&	15.55	&	-	&	2012 Sept	&	\citet{Folsom2016}	\\
HD 166435	&	1.04	&	0.99	&	3.43	&	3800	&	1.88	&	6.53	&	4.05	&	-	&	Petit et al., (in prep)	\\
HD 175726	&	1.06	&	1.06	&	3.92	&	500	&	2.40	&	4.71	&	2.11	&	-	&	Petit et al., (in prep)	\\
V 447 Lac	&	0.9	&	0.81	&	4.43	&	257	&	2.50	&	4.17	&	2.83	&	2014 Jun	&	\citet{Folsom2016}	\\
HN Peg	&	1.085	&	1.04	&	4.6	&	260	&	6.05	&	7.06	&	1.33	&	2007 Jul	&	\citet{Saikia2015}	\\
...	&	...	&	...	&	...	&	...	&	2.96	&	4.65	&	1.76	&	2008 Aug	&	...	\\
...	&	...	&	...	&	...	&	...	&	4.39	&	5.91	&	1.33	&	2009 Jun	&	...	\\
...	&	...	&	...	&	...	&	...	&	5.44	&	7.30	&	1.29	&	2010 Jul	&	...	\\
...	&	...	&	...	&	...	&	...	&	4.56	&	7.72	&	2.06	&	2011 Jul	&	...	\\
...	&	...	&	...	&	...	&	...	&	8.17	&	9.46	&	1.13	&	2013 Jul	&	...	\\
TYC 5164-567-1	&	0.9	&	0.89	&	4.68	&	120	&	22.57	&	23.51	&	-	&	2013 Jun	&	\citet{Folsom2016}	\\
HD 39587	&	1.03	&	1.05	&	4.83	&	500	&	2.75	&	6.64	&	2.18	&	-	&	Petit et al., (in prep)	\\
HIP 12545	&	0.95	&	1.07	&	4.83	&	24	&	40.75	&	45.62	&	1.06	&	2012 Sept	&	\citet{Folsom2016}	\\
HD 72905	&	1	&	1	&	5	&	500	&	1.90	&	4.57	&	3.23	&	-	&	Petit et al., (in prep)	\\
DX Leo	&	0.9	&	0.81	&	5.38	&	257	&	7.85	&	8.48	&	-	&	2014 May	&	\citet{Folsom2016}	\\
V 439 And	&	0.95	&	0.92	&	6.23	&	257	&	4.31	&	4.51	&	-	&	2014 Sept	&	\citet{Folsom2016}	\\
HD 190771	&	0.96	&	0.98	&	8.8	&	2700	&	2.84	&	3.98	&	2.09	&	-	&	Petit et al., (in prep)	\\
$\kappa$ Ceti	&	1.03	&	0.95	&	9.3	&	600	&	4.38	&	6.23	&	1.85	&	2012 Oct	&	\citet{Nascimento2016}	\\
HD 73350	&	1.04	&	0.98	&	12.3	&	510	&	1.69	&	3.44	&	2.84	&	-	&	Petit et al., (in prep)	\\
HD 73256	&	1.05	&	0.89	&	14	&	830	&	1.21	&	2.12	&	2.66	&	2008 Jan	&	\citet{Fares2013}	\\
HD 56124	&	1.03	&	1.01	&	18	&	4500	&	1.11	&	1.12	&	-	&	-	&	Petit et al., (in prep)	\\
18 Sco	&	0.98	&	1.02	&	22.7	&	4700	&	0.42	&	0.62	&	2.26	&	2007 Aug	&	\citet{Petit2008}	\\
HD 9986	&	1.02	&	1.04	&	23	&	4300	&	0.33	&	0.34	&	-	&	-	&	Petit et al., (in prep)	\\
HD 46375	&	0.97	&	0.86	&	42	&	5000	&	0.81	&	0.83	&	-	&	2008 Jan	&	\citet{Fares2013}	\\
	\hline
\end{tabular}
\small
\end{center}
$^a$We have listed the mass for AB Dor as 1$\solarmass$ but there have been a range of reported masses for AB Dor in the literature. See the footnote in section \ref{subsec:PFSS} for further discussion.
\end{minipage}
\end{table*}

\begin{figure}
	\begin{center}
	\includegraphics[trim=1cm 1cm 1cm 0cm,width=\columnwidth]{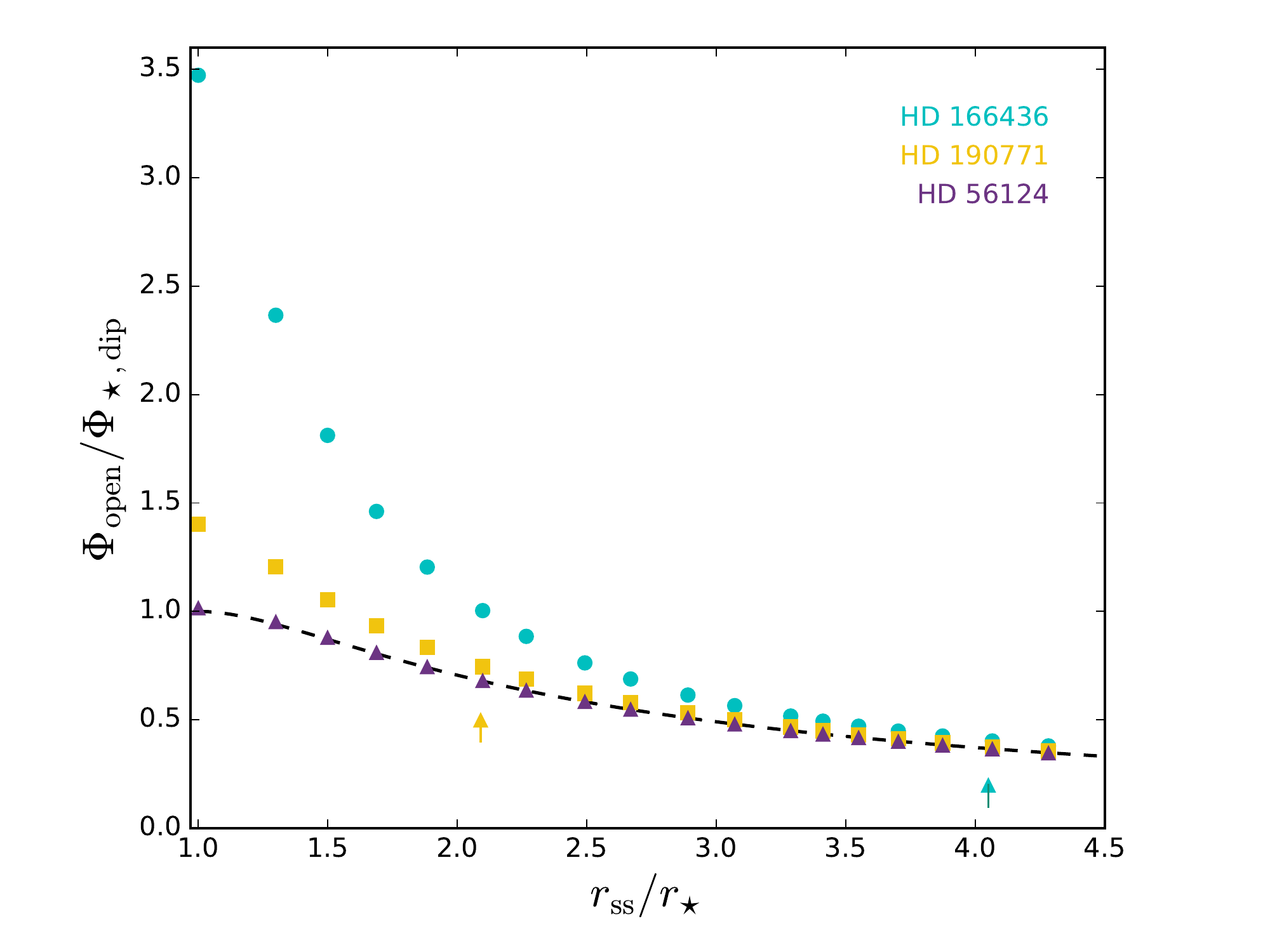}
	\end{center}
	\caption{Open flux normalised to the surface dipolar flux against source surface radii for HD 166435 (cyan circles), HD 190771 (yellow squares) and HD 56124 (purple triangles). At $r_{\rm ss}/r_{\star}=1$, the open flux is equal to flux at the stellar surface since all the magnetic flux is forced to be open. The dashed line represents the normalised open flux from a purely dipolar field, i.e. equation (\ref{eq:OpenFluxRatio}). Cyan and yellow arrows indicate the $r_{\rm ss, 10\%}$ values for HD 166435 and HD 190771. No arrow is shown for HD 56124 since the discrepancy between the open fluxes determined from the multipolar case and the dipolar case never exceed 10\%.}
	\label{fig:FullVsDipComp}
\end{figure}

\begin{table*}
\begin{minipage}{140mm}
	\begin{center}	
	\caption{For a range of source surface radii, $r_{{\rm ss,}i}$, we perform a fit of the form $\Phi_{{\rm open,}i} = 10^{c_i}P^{m_i}_{\rm rot}$ to the unsaturated stars. Here we list the $m_i$ and $c_i$ values for each $r_{{\rm ss,}i}$.} 
	\label{tab:fits}
	\begin{tabular}{lcccccc}
	\hline
	\hline
$r_{{\rm ss,}i}/r_{\star}$	&	1.00	&	1.30	&	1.50	&	1.69	&	1.88	&	2.10	\\
$m_i$	&	 -1.545 $\pm$ 0.200	&	-1.522 $\pm$ 0.206 	&	-1.525 $\pm$ 0.211 	&	-1.533 $\pm$ 0.215 	&	-1.540 $\pm$ 0.217 	&	-1.547 $\pm$ 0.219 	\\
$c_i$	&	24.96 $\pm$ 0.18	&	24.86 $\pm$ 0.19	&	24.80 $\pm$ 0.20	&	24.75 $\pm$ 0.20	&	24.71 $\pm$ 0.20	&	24.67 $\pm$ 0.21	\\
\vspace{2mm}													\\
$r_{{\rm ss,}i}/r_{\star}$	&	2.27	&	2.49	&	2.67	&	2.89	&	3.07	&	3.29	\\
$m_i$	&	-1.551 $\pm$ 0.221 	&	-1.554 $\pm$ 0.222 	&	-1.557 $\pm$ 0.222 	&	-1.559 $\pm$ 0.223 	&	-1.561 $\pm$ 0.223 	&	-1.563 $\pm$ 0.224 	\\
$c_i$	&	24.63 $\pm$ 0.21	&	24.60 $\pm$ 0.21	&	24.57 $\pm$ 0.21	&	24.53 $\pm$ 0.21	&	24.51 $\pm$ 0.21	&	24.48 $\pm$ 0.21	\\
\vspace{2mm}													\\
$r_{{\rm ss,}i}/r_{\star}$	&	3.41	&	3.55	&	3.70	&	3.87	&	4.07	&	4.28	\\
$m_i$	&	-1.564 $\pm$ 0.224 	&	-1.565 $\pm$ 0.224 	&	-1.566 $\pm$ 0.224 	&	-1.566 $\pm$ 0.224 	&	-1.567 $\pm$ 0.225 	&	-1.568 $\pm$ 0.225 	\\
$c_i$	&	24.46 $\pm$ 0.21	&	24.45 $\pm$ 0.21	&	24.43 $\pm$ 0.21	&	24.41 $\pm$ 0.21	&	24.39 $\pm$ 0.21	&	24.37 $\pm$ 0.21	\\
	\hline
\end{tabular}
\end{center}
\end{minipage}
\end{table*}

\subsection{Multipolar method of determining open flux}
\label{subsec:FullOpenFlux}
In this section, we will estimate the open flux using the full set of spherical harmonics available from the ZDI maps. Unfortunately, there is no simple analytic method of estimating the open flux from the surface flux analogous to equation (\ref{eq:OpenFluxRatio}) due to the summation over different modes in equation (\ref{eq:Br}). We must therefore numerically evaluate equation (\ref{eq:Open}) for a range of source surface radii to determine the impact of the higher order modes on the open flux. Fig. \ref{fig:FullVsDipComp} shows the open flux, $\Phi_{\rm open}$, normalised to the surface dipolar flux, $\Phi_{\rm \star,dip}$, against source surface radius for three stars from our sample. A dashed line represents the open flux expected in a purely dipolar case, i.e. as calculated by equation (\ref{eq:OpenFluxRatio}). The three chosen stars represent three cases. The surface magnetic field of HD 56124  (purple triangles), as determined from ZDI, is strongly dipolar. As such, it follows the pure dipole case (dashed line) very closely. On the other hand, the dipolar component represents only a small fraction of the magnetic energy in the ZDI map of HD 166435 (cyan circles). This star therefore shows large deviations from the pure dipole case at the smallest values of $r_{\rm ss}$. HD 190771 (yellow squares) represents an intermediate case and is also representative of the majority of the stars in our sample. From Fig. \ref{fig:FullVsDipComp}, it is clear to see why \citet{See2017} found that the dipole component dominated the open flux for their chosen source surface radii of $r_{\rm ss} = 3.41r_{\star}$. Even for HD 166435, which has one of the weakest dipole components in our sample, the effects of higher order spherical harmonics have become small by $r_{\rm ss} = 3.41r_{\star}$. In table \ref{tab:ZDIStars}, we list $r_{\rm ss, 10\%}$ values for each star. This is the source surface radius at which the discrepancy between the open flux calculated using the full set of spherical harmonic modes and the open flux calculated considering just the dipolar mode exceeds 10\% (no value is listed if it never exceeds 10\%). We also indicate the $r_{\rm ss, 10\%}$ values for HD 166435 and HD 190771 with cyan and yellow arrows in Fig. \ref{fig:FullVsDipComp}. The choice of 10\% is arbitrary but serves to illustrate how dominant the dipolar mode is for each star. In most cases, the discrepancy between the two methods only exceeds 10\% at relatively small $r_{\rm ss}$ values; less than $2r_{\star}$ for the majority of our sample and less than 3$r_{\star}$ for all but two stars.

\begin{figure}
	\begin{center}
	\includegraphics[trim=1cm 1cm 1cm 0cm,width=\columnwidth]{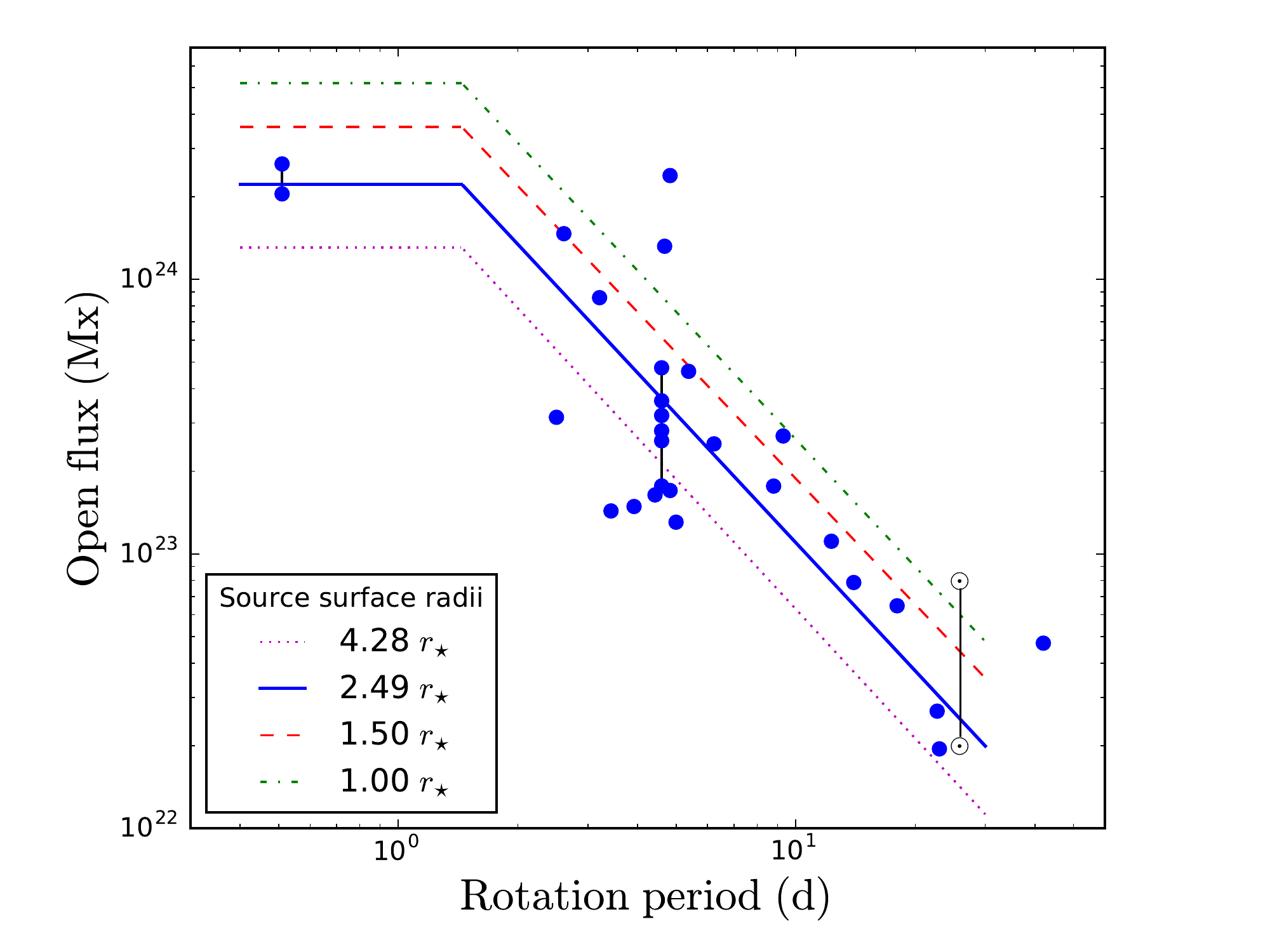}
	\end{center}
	\caption{Open flux against rotation period calculated using $r_{\rm ss}=2.49r_{\star}$ (blue circles). Stars observed at multiple epochs are connected by a vertical line. A fit to these points is shown with a solid blue line. Fits to the open flux calculated using $r_{\rm ss} = 1.00r_{\star},\ 1.50r_{\star}\ \&\ 4.28r_{\star}$ are also shown. The data points for these fits are translated vertically in the plot when compared to the $r_{\rm ss}=2.49r_{\star}$ data points and show a similar level of dispersion. However, they are not plotted for clarity. The coefficients for these fits, as well as other fits using different $r_{\rm ss}$ values, can be found in table \ref{tab:fits}. Two solar symbols indicate the variation of the open flux over the solar cycle.}
	\label{fig:FullOpenFlux}
\end{figure}

In order to determine the effect of using the full spherical harmonic decomposition available in the ZDI maps rather than just the dipole components, we require a method of estimating the open flux as a function of rotation period and source surface radius. This is done as follows. First we calculate the open flux of all the stars in our sample for a discrete set of different source surface radii, $r_{{\rm ss,}i}$, in the range $1r_{\star}$ to $4.28r_{\star}$. Here, $i$ is an index labeling the different source surface radii in this discrete set. For each source surface radii, $r_{{\rm ss,}i}$, we perform a fit to the unsaturated stars, similarly to the one undertaken in section \ref{subsec:DipOpenFlux}, of the form $\Phi_{{\rm open,}i} = 10^{c_i}P^{m_i}_{\rm rot}$, where $m_i$ and $c_i$ are fit parameters. As in section \ref{subsec:DipOpenFlux}, the saturated value of the open flux is taken to be $10^{c_i}P^{m_i}_{\rm rot,crit}$. In Fig. \ref{fig:FullOpenFlux}, we show the open flux values for one choice of the source surface radii, $r_{{\rm ss,}i}=2.49r_{\star}$ (blue points) as well as the fit to these points (solid blue line). This is the typical source surface radii chosen when studying the Sun. The PFSS model predicts solar open fluxes between $\sim 2 \times 10^{22}\ {\rm Mx}$ and $\sim 8 \times 10^{22}\ {\rm Mx}$ \citep{Jardine2017}. For a solar rotation period, we predict an open flux of $2.5 \times 10^{22}\ {\rm Mx}$ using our fit, in good agreement with the solar values. Additionally, we also show the fits for three other choices of $r_{{\rm ss,}i}$ to illustrate the impact of $r_{{\rm ss,}i}$ on the fits. The data points for these fits are not shown for clarity. A full list of $m_i$ and $c_i$ coefficients for all values of $r_{{\rm ss,}i}$ are available in table \ref{tab:fits}. 

To calculate the open flux for an arbitrary $P_{\rm rot}$ and $r_{\rm ss}$, we pick the two $r_{{\rm ss},i}$ values from table \ref{tab:fits} that bound our choice of $r_{\rm ss}$. Using the $m_i$ and $c_i$ values associated with these two $r_{{\rm ss},i}$ values, we calculate two open flux values at the rotation period of interest, i.e. $\Phi_{{\rm open,}i} = 10^{c_i}P^{m_i}_{\rm rot}$. Lastly, we interpolate between these two $\Phi_{{\rm open,}i}$ values to determine the open flux for our chosen $r_{\rm ss}$. When calculating the open flux using a $r_{\rm ss}>4.28r_{\star}$, the dipolar term strongly dominates (see Fig. \ref{fig:FullVsDipComp}) and we simply use equation (\ref{eq:OpenFluxRatio}).

\begin{table*}
\begin{minipage}{105mm}
	\begin{center}	
	\caption{The open cluster data used in this study. For each cluster, we list the age as well as the 25th, 50th and 90th percentile of the angular velocity distribution of $\sim$solar-mass stars.} 
	\label{tab:cluster}
	\begin{tabular}{lccccc}
	\hline
	Cluster & Age & $\Omega_{25}$ & $\Omega_{50}$ & $\Omega_{90}$ & Ref. \\
	name & (Myr) & ($\Omega_{\odot}$) & ($\Omega_{\odot}$) & ($\Omega_{\odot}$) & \\
	\hline
Pleiades	&	125	&	4.83$\pm$0.14	&	6.14$\pm$0.24	&	44.91$\pm$9.63	&	\citet{Hartman2010}	\\
M50	&	130	&	4.39$\pm$0.46	&	5.45$\pm$0.34	&	23.01$\pm$8.67	&	\citet{Irwin2009}	\\
M35	&	150	&	4.39$\pm$0.10	&	5.12$\pm$0.23	&	24.90$\pm$4.90	&	\citet{Meibom2009}	\\
M34	&	220	&	3.77$\pm$0.17	&	4.75$\pm$0.63	&	28.89$\pm$4.74	&	\citet{Meibom2011M34}	\\
M37	&	550	&	3.08$\pm$0.04	&	3.34$\pm$0.03	&	4.46$\pm$0.67	&	\citet{Hartman2009}	\\
Praesepe	&	580	&	2.64$\pm$0.05	&	2.76$\pm$0.04	&	2.85$\pm$0.03	&	\citet{Delorme2011}	\\
Hyades	&	625	&	2.68$\pm$0.07	&	2.75$\pm$0.06	&	3.07$\pm$0.05	&	\citet{Delorme2011}	\\
NGC 6811	&	1000	&	2.32$\pm$0.02	&	2.42$\pm$0.02	&	2.59$\pm$0.05	&	\citet{Meibom2011}	\\
NGC6819	&	2500	&	1.20$\pm$0.01	&	1.22$\pm$0.01	&	1.39$\pm$0.05	&	\citet{Meibom2015}	\\
M67	&	4200	&	0.83$\pm$0.02	&	0.92$\pm$0.03	&	1.04$\pm$0.03	&	\citet{Barnes2016}	\\
	\hline
\end{tabular}
\end{center}
\end{minipage}
\end{table*}

\begin{figure}
	\begin{center}
	\includegraphics[trim=1cm 1cm 1cm 0cm,width=\columnwidth]{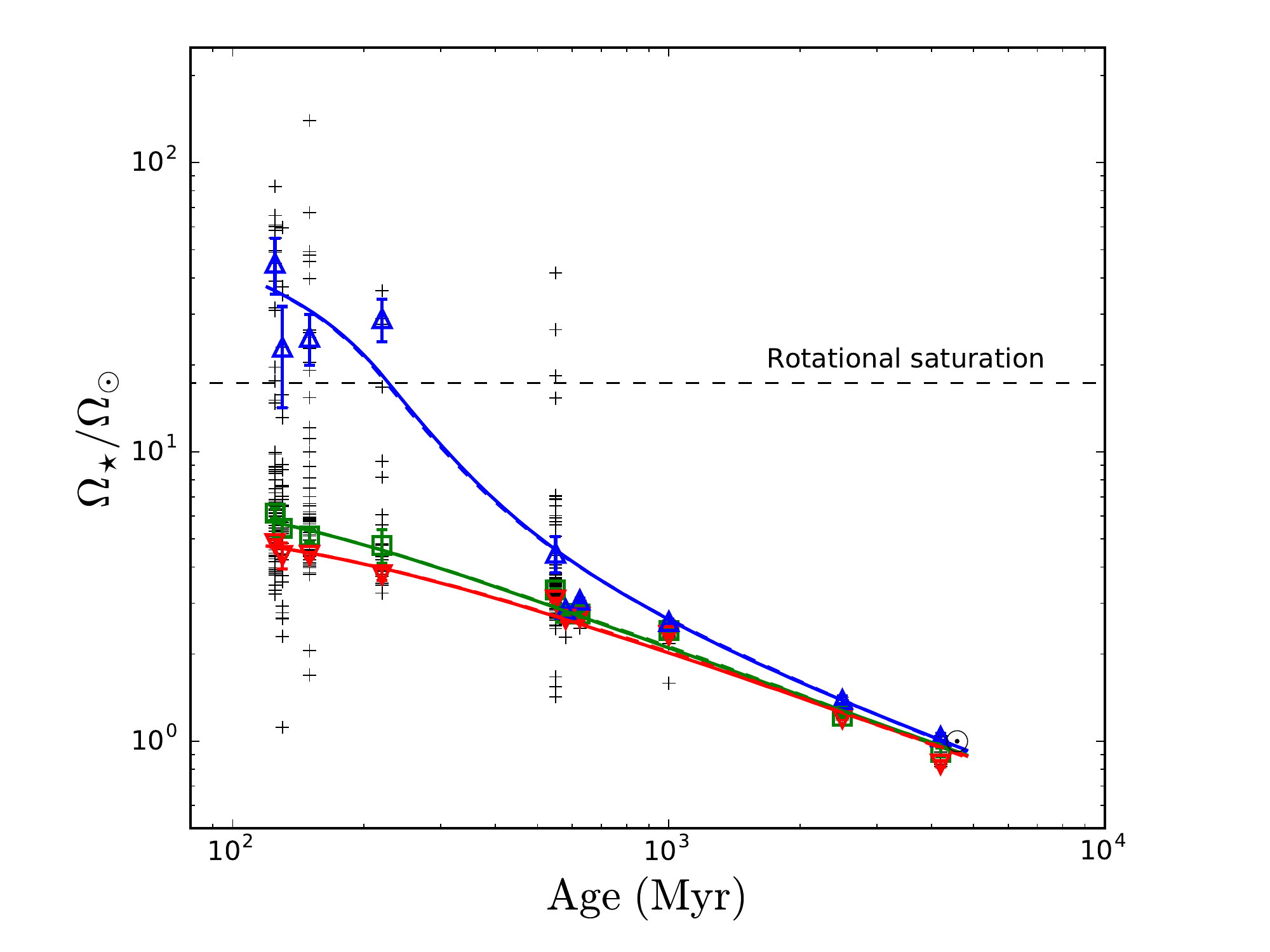}
	\end{center}
	\caption{The rotation evolution of a solar-mass star. Plus symbols indicate the observed rotation periods of solar-mass stars in open clusters. Blue upwards triangles, green squares and red downwards triangles represent the 90th, 50th and 25th percentile in each of the clusters. The blue, green and red tracks show the rotation evolution of a fast, intermediate and slow solar-mass star as calculated with the dipole method (solid lines) and the multipolar method (dashed lines). The horizontal dashed line indicates the saturation threshold. Data and references for the cluster data can be found in table \ref{tab:cluster}.}
	\label{fig:Tracks}
\end{figure}

\section{Open cluster data}
\label{sec:Clusters}
In order to constrain our model, we use the rotation periods of stars in open clusters of known ages. Since we are studying rotation period evolution on the main sequence, we have chosen clusters that have ages of 125 Myr or greater. In recent years, rotation period measurements in clusters older than 1 Gyr have been possible thanks to the Kepler Space Telescope \citep{Meibom2015,Barnes2016}. In particular, rotation period measurements from the 4 Gyr cluster M67 confirm that the Sun has a typical rotation period for a star of its age and mass \citep{Barnes2016}. The clusters we have used in this work and their ages are listed in table \ref{tab:cluster}. For each cluster, we will consider all the stars with masses between $0.9\solarmass$ - $1.1\solarmass$ to be representative of solar-mass stars. Fig. \ref{fig:Tracks} shows the distribution of rotation periods for these stars in each of the clusters as a function of age (plotted with grey plus symbols) as well as our rotational evolution tracks (these will be discussed in section \ref{sec:Tracks}). We can see that the rotation period distributions evolve with time. At early ages ($<200$ Myr), solar-mass stars can have a large range of rotation periods with the fastest spinning nearly 100 times faster than the Sun. However, after $\sim$1 Gyr, the rotation periods have nearly all converged onto a single valued track regardless of their rotational history. Similar to previous studies \citep{Gallet2013,Gallet2015,Johnstone2015}, we will fit our model to the 25th, 50th and 90th percentiles of the rotation period distributions in each cluster. Implicit in this method is the assumption that a star at a given percentile will remain at that percentile throughout its entire evolution. We use a boot strapping method to determine the rotation period at these percentiles for each cluster, as well as their errors. These are listed for each cluster in table \ref{tab:cluster}. They are also plotted with red downwards triangles (25th percentile), green squares (50th percentile) and blue upwards triangles (90th percentile) in Fig. \ref{fig:Tracks}.

\section{The rotation evolution of a solar-mass star}
\label{sec:Tracks}
In this section, we will fit rotation evolution tracks to the 90th, 50th and 25th percentiles in each of the open clusters. We will refer to these as the fast, intermediate and slow tracks respectively. In order to determine the best fit values for the power law index of the source surface radii scaling, $n$, and the scaling constant used to determine the spin-down torque, $k$, we require a goodness of fit parameter. We will use 

\begin{equation}
	X = \sum_{j} (\log\Omega_{{\rm obs,}j}-\log\Omega_{{\rm model,}j})^2.
	\label{eq:chi2}
\end{equation}
Here, $\Omega_{{\rm obs,}j}$ refers to the observed angular velocities from open clusters and $\Omega_{{\rm model,}j}$ refers our model's estimate of $\Omega_{{\rm obs,}j}$. The summation over the index, $j$, is performed over the 25th, 50th and 90th percentiles for every cluster as shown in table \ref{tab:cluster}. This is a similar goodness of fit parameter to that used by \citet{Johnstone2015}. However, unlike these authors, we do not assign different weights to the different clusters. Tests indicate that giving older clusters a larger weighting does not significantly change our results.

\begin{figure}
	\begin{center}
	\includegraphics[trim=1cm 1cm 1cm 0cm,width=\columnwidth]{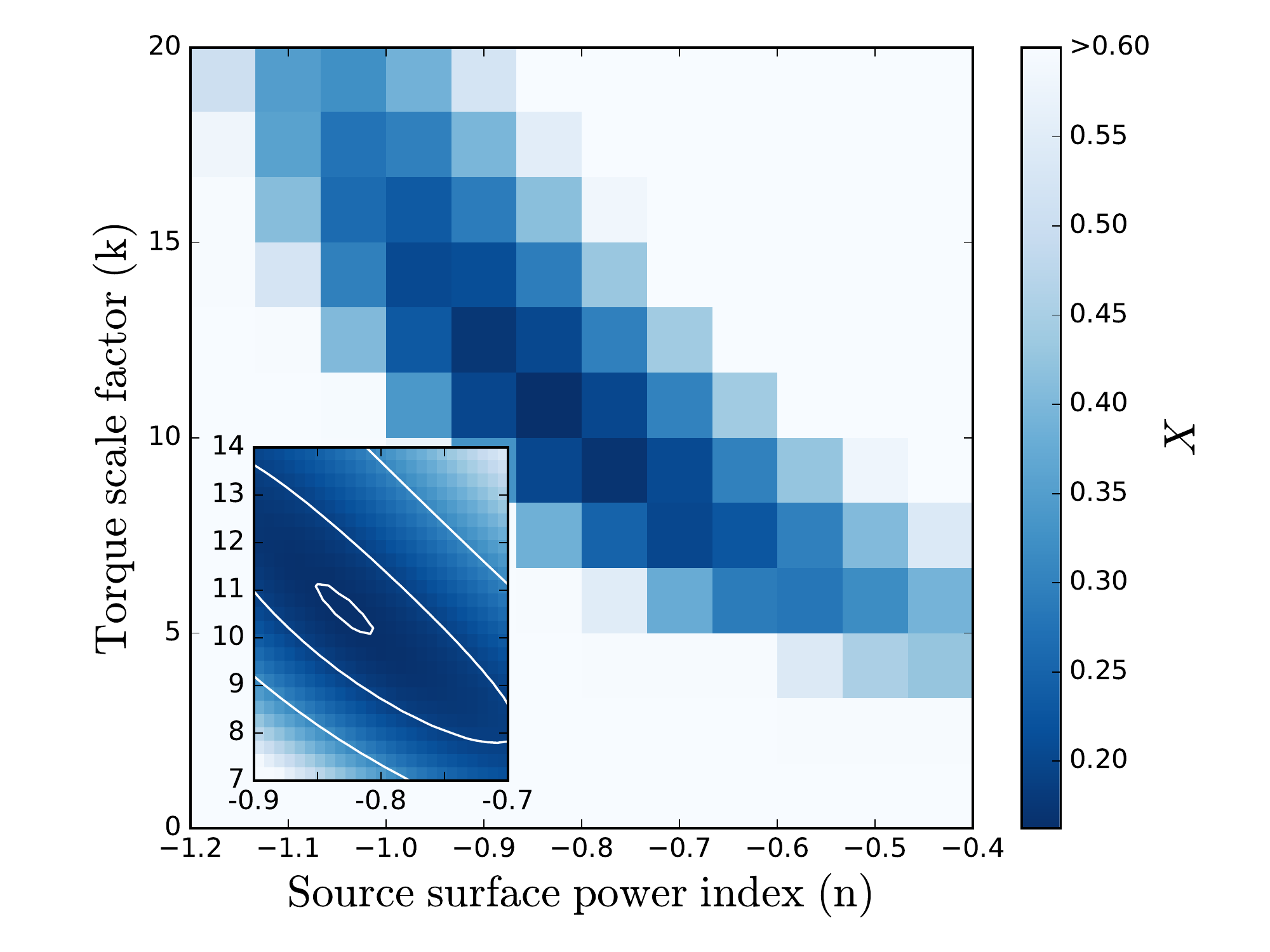}
	\end{center}
	\caption{The goodness-of-fit parameter, $X$, calculated over a grid of source surface power indices, $n$, and torque scale factors, $k$, for the dipolar method. $X$ values larger than 0.6 have been truncated to 0.6. The inset shows a higher resolution search through the $n$ and $k$ values around the minima. Contours for $X=\{0.16,0.19,0.3\}$ are also shown on the inset. The minimum in $X$ occurs at $n_{\rm dip}=-0.84$, $k_{\rm dip}=10.64$.}
	\label{fig:chi2}
\end{figure}

In Fig. \ref{fig:chi2}, we calculate the value of $X$ over a grid of $n$ and $k$ values using the dipole method of determining the open flux (as described in section \ref{subsec:DipOpenFlux}). There is a well defined minimum in $X$ that occurs at $n_{\rm dip}=-0.84$ and $k_{\rm dip}=10.64$. It is worth noting that some degeneracy exists between $n_{\rm dip}$ and $k_{\rm dip}$ however. A more (less) negative value of $n_{\rm dip}$ can be partially offset by a larger (smaller) $k_{\rm dip}$ value for only a small increase in our goodness-of-fit parameter. Using $n_{\rm dip}$ and $k_{\rm dip}$, we calculate the fast, intermediate and slow tracks which are plotted with solid red, green and blue lines in Fig. \ref{fig:Tracks}. By combining equations (\ref{eq:OpenFluxRatio}), (\ref{eq:rss}) \& (\ref{eq:rss2}) with the fit from Fig. \ref{fig:SurDipFlux} and the value of $n_{\rm dip}$, we can also determine the functional dependence of the open flux on the rotation period in our model. This is given by

\begin{equation}
	\Phi_{\rm open,dip} = 7.27\times 10^{23} \frac{\tilde{P}_{\rm rot}^{-3.26}}{31.3\tilde{P}_{\rm rot}^{-2.52}+1}, \hspace{3mm} (P_{\rm rot}>P_{\rm rot,crit})
	\label{eq:openFluxPeriod}
\end{equation}
and
\begin{equation}
	\Phi_{\rm open,dip,sat} = 2.00\times 10^{23}, \hspace{3mm} (P_{\rm rot}<P_{\rm rot,crit}),
	\label{eq:openFluxPeriod2}
\end{equation}
where $\tilde{P}_{\rm rot}=P_{\rm rot}/P_{\rm rot,\odot}$.

The fact that $k_{\rm dip}=10.64$ has a value that is greater than 1 suggests that we are underestimating the spin-down torque in some way. This could be attributed to a number of reasons. For instance, the coronal temperature, a parameter which is kept fixed in the simulations of \citet{Reville2015} can affect the rate at which angular momentum is lost from a star, even for a fixed mass-loss rate (Pantolmos et al., in prep). Additionally, the mass-loss rates we use may be systematically underestimated causing a lower than expected spin-down torque. It is worth emphasising that mass-loss rates are extremely difficult to estimate and poorly constrained observationally. Differences between the mass-loss rate estimated by the model of \citet{Cranmer2011} and the real mass-loss rates of stars could be absorbed into the fit parameter $k$ (and perhaps $n$). Currently, there is no way of determining how much of the fact that $k_{\rm dip}$ is larger than 1 can be attributed to inaccurately estimating mass-loss rates. Although the mass-loss rates predicted by the model of \citet{Cranmer2011} agree reasonably well with observations of the Sun, they may be less accurate for younger or more rapidly rotating stars. Lastly, we have assumed solid body rotation for simplicity. However, if core-envelope decoupling were included, the stellar wind braking should be more efficient since it would be acting on the envelope only. This should reduce the value of $k_{\rm dip}$ that we obtain with our model. \citet{Johnstone2015} obtain torque scaling value of 11 within their model which is comparable to our value. However, \citet{Gallet2015} obtained a value of 1.7 in their solar mass model which suggests their models may be capturing the relevant physics more accurately.

It is also worth noting that the particular values of $n_{\rm dip}$ and $k_{\rm dip}$ we obtain are dependent on the form of the flux-rotation relation that we adopt, i.e. the fit from Fig. \ref{fig:SurDipFlux}. For example, if we only fit to the stars in Fig. \ref{fig:SurDipFlux} with masses $0.95\solarmass \leq M_{\star} \leq 1.05\solarmass$, we recover $n_{\rm dip}=-0.67$ and $k_{\rm dip}=8.25$. It is clear that more ZDI observations of solar-mass mass stars are needed in order to refine our model. For the remainder of this work we will consider the canonical $n_{\rm dip}$ and $k_{\rm dip}$ values to be those determined using the full ZDI sample, i.e. the stars with masses $0.9\solarmass \leq M_{\star} \leq 1.1\solarmass$ since this matches the mass bin width chosen for the open cluster rotation period data.

We perform this procedure again but using the multipolar method to determine the open flux (as described in section \ref{subsec:FullOpenFlux}). The equivalent plot of Fig. \ref{fig:chi2} for the multipolar method looks very similar (not shown) with a minimum in $X$ occurring at $n_{\rm multi}=-0.82$ and $k_{\rm multi}=10.18$. These values are similar to those calculated using the dipole only method. Using the multipolar method of determining the open flux in conjunction with the $n_{\rm multi}$ and $k_{\rm multi}$ values, we plot the fast, intermediate and slow rotator tracks in Fig. \ref{fig:Tracks} with dashed lines. The dashed rotation tracks lay almost exactly on top of the solid rotation tracks determined using the dipolar method.

\begin{figure}
	\begin{center}
	\includegraphics[trim=1cm 1cm 1cm 0cm,width=\columnwidth]{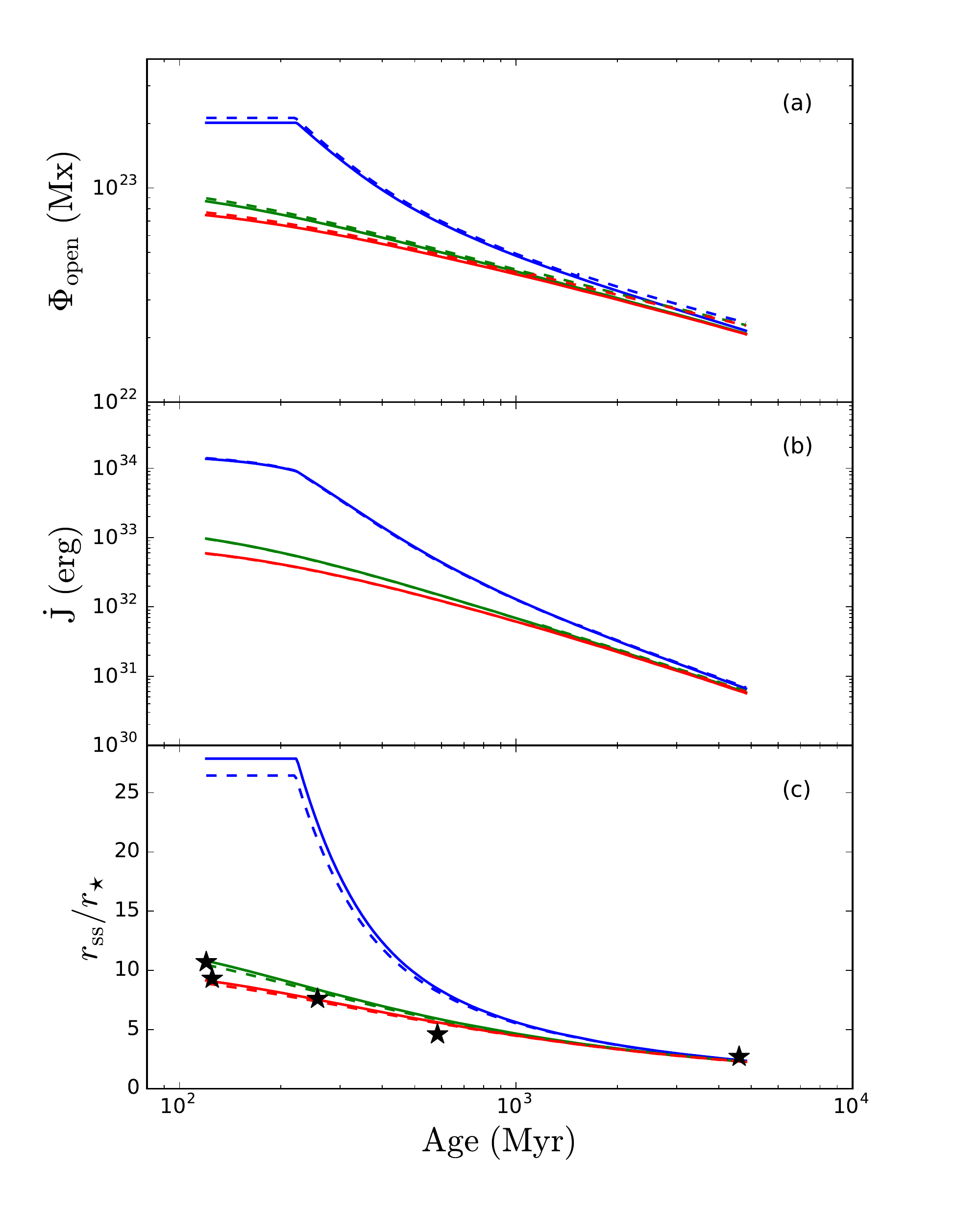}
	\end{center}
	\caption{(a) Open flux, (b) angular-momentum-loss rate and (c) source surface radii against age. In each panel, tracks are shown for fast (blue), intermediate (green) and slow (red) rotators calculated using the dipolar (solid lines) and multipolar (dashed lines) methods. In panel (c), the optimal source surface radii for a number of roughly solar-mass stars, as determined by \citet{Reville2016}, are shown with star symbols (see text).}
	\label{fig:agePlot}
\end{figure}

In Figs. \ref{fig:agePlot}a and \ref{fig:agePlot}b, we plot the open flux and angular momentum-loss rate for both the dipolar and multipolar methods. The open flux and angular momentum-loss rates are both monotonically decreasing functions of age for the fast, intermediate and slow tracks. However, the fast tracks show a change in behaviour at $\sim$200 Myr. As can be seen in Fig. \ref{fig:Tracks}, this is the age at which fast track stars transition from the saturated to the unsaturated regime. Unlike the fast track, stars on the intermediate and slow tracks are never rotating quickly enough to be in the saturated regime.

To understand why the two methods produce such similar results, we need to understand how the source surface radii evolves over the main sequence. Fig. \ref{fig:agePlot}c shows the source surface radius evolution for the fast, intermediate and slow tracks. These are calculated using equations (\ref{eq:rss}) and (\ref{eq:rss2}) in conjunction with the rotation evolution tracks shown in Fig. \ref{fig:Tracks} and our best fit values for $n_{\rm dip}$ and $n_{\rm multi}$. Over the course of the main sequence, the source surface radii of these stars shrink as the star spins down and magnetic activity declines. For the intermediate and slow rotators, the source surface radius steadily drops from $\sim 10r_{\star}$ at $\sim 100\ {\rm Myr}$ to $\sim 2.5 r_{\star}$ by the age of the Sun. However, the fast rotator is spinning rapidly enough during its early main sequence lifetime that its source surface radius attains the saturation value of $\sim 25 r_{\star}$. Indeed, no solar-mass star can have a source surface radius larger than the saturation value under this model. From Fig. \ref{fig:FullVsDipComp} and table \ref{tab:ZDIStars}, we see that, for the majority of the stars in our sample, large differences in the open flux obtained from the dipolar and multipolar methods only occur at source surface radii smaller than roughly two or three stellar radii. Since all three tracks (and the tracks for any other percentile one might calculate) maintain source surface radii larger than $2.5 r_{\star}$ until the age of the Sun, it should be no surprise that there is very little difference between the dipolar and multipolar methods for calculating the open flux in the range of ages we have studied here. If we were to extend the rotation tracks to ages significantly past the age of the Sun, it is possible that the source surface radii would become small enough for the differences between the two methods of calculating open flux to matter. However, currently, the rotation period behaviour of stars much older than the Sun is unclear. We will discuss this issue further in section \ref{sec:Discussion}.

In Fig. \ref{fig:agePlot}c, we also plot the optimum source surface for a number of stars as estimated by \citep{Reville2016}. These authors used 3D MHD simulations to study a sample of stars that are all roughly solar-mass, using ZDI maps as a boundary condition for the magnetic field at the stellar surface. Their sample consisted of the stars BD-16351, TYC 5164-567-1, HII 296, DX Leo and AV 2177. The majority of these stars are also in our own sample. From their MHD simulations, they were able to determine the open flux of each star. They then modelled each star with a PFSS model and determined the source surface radius that would be required for the PFSS model to produce the same open flux as their MHD models. These optimum source surface radii are plotted in Fig. \ref{fig:agePlot}c with star symbols. We can compare the optimum source surface radii of \citet{Reville2016} to the source surface radii as estimated by our own model. Taking $n_{\rm multi}=-0.82$ from our multipolar method, we predict source surface radii of $13.5r_{\star}$, $10.1r_{\star}$, $16.4r_{\star}$, $9.0r_{\star}$ and $6.3r_{\star}$ for BD-16351, TYC 5164-567-1, HII 296, DX Leo and AV 2177 respectively. The optimum source surface radii that \citet{Reville2016} predict are $8.1r_{\star}$, $10.7r_{\star}$, $9.3r_{\star}$, $7.6r_{\star}$ and $4.6r_{\star}$. We see that our estimates are close to those of \citet{Reville2016} but our model tends to produce source surface radii that are larger, sometimes by a factor of $\sim 1.7$. It is worth pointing out that the stars modelled by \citet{Reville2016} are all slow/moderate rotators. Our model predicts that the source surface radii of slowly/moderately rotating stars only change by a factor of a few. A similar study to that of \citet{Reville2016} modelling stars on the rapidly rotating track, where the source surface radii appears to change much more drastically over the main sequence in our model, would provide a much more stringent comparison.

Lastly, it is worth commenting on the source surface radii values. Our model estimates that the fastest rotators have $r_{\rm ss}>25r_{\star}$ which is an order of magnitude larger than the Sun's source surface radii. Our simplified model requires that the source surface radii be this large for the fastest rotators but it is not clear whether, in reality, closed field loops can be maintained out to such a large distance. For instance, magnetocentrifugal forces acting on the coronal plasma may cause closed loops to open up closer to the stellar surface than they otherwise would have. Since our model does not self-consistently model the interactions between the stellar wind plasma and magnetic field, it is not possible to say how large an effect this might have. Such questions are left to future investigations and will require more sophisticated modelling of the relevant physics to answer.

\section{Discussion and conclusions}
\label{sec:Discussion}
In this work we have used a potential field source surface model (PFSS), in conjunction with a sample of ZDI maps, to analyse how the open flux of solar-mass stars varies as a function of rotation and source surface radius. We then use these open flux relationships and the braking law of \citet{Reville2015} to model the rotation period evolution of solar-mass stars on the main sequence up to the age of the Sun. We have assumed solid body rotation for simplicity. Within the PFSS model, the source surface radius is a free parameter. Previous works using this model have typically set the source surface radius to a value close to the solar value ($\sim 2.5r_{\star}$). However, in this work, using rotation period data from open clusters, we are able to constrain how the source surface radii varies with the rotation period of the star. We predict that the fastest rotators begin life on the main sequence with a source surface radius of $\sim 26 r_{\star}$ while the intermediate and slower rotators start out with source surface radii of $\sim 10 r_{\star}$. Eventually, the source surface radii of solar-mass stars will converge and reach the solar source surface radius by the age of the Sun.

Previous rotation period evolution models have typically used braking laws that are formulated in terms of the dipolar surface magnetic field strength \citep[e.g.][]{Matt2012}. However, we use the braking law of \citet{Reville2015}, which is formulated in terms of the open flux. In principle, this braking law allows us to account for higher order spherical harmonic modes in the surface magnetic field. In practice however, we find that the dipole component of the magnetic field dominates the open flux for all but the smallest choices of the source surface radius. As outlined in section \ref{sec:Tracks}, the dipolar open flux in our model can be analytically calculated as a function of rotation period by combining equations (\ref{eq:OpenFluxRatio}), (\ref{eq:rss}) and (\ref{eq:rss2}) with the fit from Fig. \ref{fig:SurDipFlux}. 

When considering the effect of field geometry on angular momentum evolution, our results suggest that it would be reasonable to use the braking law of \citet{Matt2012} over that of \citet{Reville2015}. However, some caution should be exercised when directly comparing models using the two braking laws. The model we have presented uses two fit parameters. These are the power law index for the source surface radius, $n$, and the torque scaling parameter, $k$. Other models that use the braking law of \citet{Matt2012} usually have a similar torque scaling parameter to the one used in this work. However, since they do not have to model the open flux, they do not need a fit parameter like our $n$. Instead these models have other free parameters. For instance, disk lifetimes are a free parameter in the work of \cite{Gallet2013} and \citet{Gallet2015} while the mass-loss rates used by \citet{Johnstone2015} are specified as power laws of mass and rotation, where the power law indices are fit parameters. Uncertainties in different models are therefore absorbed in different places making a direct comparison between models difficult. In the future, these uncertainties can be reduced through further observations of the rotation period distributions of open clusters, magnetic field strengths and disk lifetimes.

For this work, we have restricted ourselves to studying solar-mass stars on the main sequence up the age of Sun. However, this is only one part of the lifetime of a star. In comparison to the main sequence, modelling the rotation period evolution of stars on the pre-main sequence (PMS) is much more difficult. Early on, in the classical T Tauri phase, the presence of a circumstellar disc is an additional element that must be considered. Throughout the entire PMS stars spin up as they contract towards the main sequence. However, their rotational velocities are much slower than expected from contraction alone \citep{Vogel1981} indicating that significant spin-down torques are acting on PMS stars. There is strong evidence that the presence of discs inhibits the spin-up of these stars \citep{Edwards1993,Bouvier1997,Rebull2004} although the precise mechanism by which this is achieved is still unclear. Common suggestions include disc-locking \citep{Choi1996} or accretion powered stellar winds \citep{Matt2010,Matt2012PMS}. Additionally, changes in the surface magnetic field associated with internal structure changes may also play a role \citep{Gregory2012,Folsom2016}. 

In contrast to young main sequence stars, the rotation period evolution of stars older than the age of the Sun remain relatively unconstrained. Consequently rotation evolution models, such as the one we have presented, cannot be extended beyond the age of the Sun with any reliability. However, recent advances have allowed for the determination of rotation periods and asteroseismic ages of old field stars \citep{Garcia2014}. These stars appear to be rotating much faster than expected from gyrochronology. Indeed, dramatically reduced braking appears to be required to explain the rapid rotation of these stars \citep{vanSaders2016}. One possible explanation for such a reduction is that the nature of the dynamo changes at a Rossby number of $\sim 2$ such that the surface magnetic field is concentrated into smaller scales \citep{Metcalfe2016}. Under this interpretation, the Sun (Rossby number $\sim 2$) is on the verge of transitioning to a state of reduced braking. If this suggestion is true, we might expect our result, that it is the dipole component of the magnetic field that predominantly governs the rotation evolution of main sequence solar-mass stars, to break down at old ages. This suggestion will take time to confirm however, since there are currently no ZDI observations of stars with Rossby numbers much bigger than 2.

In principle, the technique we have outlined in this work can also be applied to stars of other masses. However, there would be a number of additional barriers to overcome. Firstly, more ZDI maps of saturated stars would be required. Presently, the open flux behaviour for solar-type stars in the saturated regime is relatively unconstrained in comparison to the unsaturated regime (see Fig. 2b of \citet{See2017}). As discussed in section \ref{sec:Tracks}, for solar-mass stars, only the most rapid rotators spend any time in the saturated regime and those that do, rapidly spin-down into the unsaturated regime. Therefore, the loose constraints on the saturated level of open flux for solar-mass stars is not a large problem. However, the critical rotation period at which saturation sets in increases for lower mass stars. This is easily seen in fig. 6 of \citet{Johnstone2015}. Consequently, the loose constraints in the saturated regime is more problematic for lower mass stars since they can spend more time in the saturated regime. Secondly, the method we have used to estimate our source surface radii is calibrated to the Sun (equations (\ref{eq:rss}) and (\ref{eq:rss2})). Since the source surface radii of other stars are unknown, we would have nothing to calibrate to when studying the angular momentum evolution of stars with different masses. A possible method of overcoming this problem would be to recast equation (\ref{eq:rss}) as $r_{\rm ss} = \alpha P_{\rm rot}^n$, where $\alpha$ is a constant of proportionality. This would, however, introduce another parameter to fit for. Finally, the magnetic properties of the very lowest masses ($<0.2\solarmass$) appear to exist in one two states; either strong and dipolar or weak and multipolar \citep{Morin2010}. \citet{See2017} showed that the spin-down properties corresponding to these two states are very different with the instantaneous spin-down time-scales of the strong dipolar stars being two orders of magnitude shorter than their weak multipolar counterparts. As discussed by these authors, detailed angular momentum evolution modeling of these stars must wait until the number of $<0.2\solarmass$ stars of known ages that have been mapped with ZDI is vastly expanded.

\section*{Acknowledgements}
The authors thank an anonymous referee that helped improve the quality of this manuscript as well as Sean Matt, Florian Gallet and Isabelle Baraffe for useful discussions regarding this work. VS acknowledges support from a Science \& Technology Facilities Council (STFC) postdoctoral fellowship and the European Research Council Consolidator grant AWESoMeStars. SBS and SVJ acknowledge research funding by the Deutsche Forchungsgemeinschaft (DFG) under grant SFB, project A16. RF acknowledge financial support by WOW from INAF through the Progetti Premiali funding scheme of the Italian Ministry of Education, University, and Research. This study was supported by the grant ANR 2011 Blanc SIMI5-6 020 01 ``Toupies: Towards understanding the spin evolution of stars" (\url{http://ipag.osug.fr/Anr_Toupies/}). This work is based on observations obtained with ESPaDOnS at the CFHT and with NARVAL at the TBL. CFHT/ESPaDOnS are operated by the National Research Council of Canada, the Institut National des Sciences de l'Univers of the Centre National de la Recherche Scientifique (INSU/CNRS) of France and the University of Hawaii, while TBL/NARVAL are operated by INSU/CNRS. We thank the CFHT and TBL staff for their help during the observations.

\bibliographystyle{mnras}
\bibliography{RotEvo}{}

\appendix
\section{Deriving the dipolar open flux}
\label{app:Dipole}
In this appendix, we derive the ratio of open flux to surface flux for a pure dipole mode. We remind the reader that the radial component of the magnetic field, $B_{\rm r}$, in the PFSS model is given by 

\begin{equation}
	B_{\rm r}=-\sum\limits_{l=1}^N \sum\limits_{m=-l}^l [la_{lm} r^{l-1} - \left(l+1\right) b_{lm} r^{-\left(l+2\right)}] P_{lm} \left(\cos \theta \right) e^{im\phi}.
	\label{eqApp:Br}
\end{equation}
From equations (\ref{eq:Btheta}) \& (\ref{eq:Bphi}), the condition that $B_{\theta}(r_{\rm ss})=B_{\phi}(r_{\rm ss})=0$ requires that the $a_{lm}$ and $b_{lm}$ coefficients obey the relation

\begin{equation}
	a_{lm}r_{\rm ss}^{l-1} + b_{lm}r_{\rm ss}^{-(l+2)}=0.
	\label{eqApp:Constraint}
\end{equation}
Combining equations (\ref{eqApp:Br}) and (\ref{eqApp:Constraint}), one finds that

\begin{equation}
	B_{\rm r} = \sum\limits_{l=1}^N \sum\limits_{m=-l}^l B_{lm} f_l(r) P_{lm} e^{im\phi},
	\label{eqApp:ReducedBr}
\end{equation}
where $B_{lm}$ and $f_{l}(r)$ is given by

\begin{equation}
	B_{lm} = -a_{lm}lr_{\star}^{l-1} + b_{lm}(l+1)r_{\star}^{-(l+2)},
	\label{eqApp:Blm}
\end{equation}

\begin{equation}
	f_{l}(r)=\left[	\frac{(l+1)\tilde{r}^{-(l+2)}+l\tilde{r}_{\rm ss}^{-(2l+1)}\tilde{r}^{l-1}}{l\tilde{r}_{\rm ss}^{-(2l+1)}+(l+1)}\right],
	\label{eqApp:f}
\end{equation}
where $\tilde{r}= r/r_{\star}$ and $\tilde{r}_{\rm ss} = r_{\rm ss}/r_{\star}$. For a dipole, equation (\ref{eqApp:ReducedBr}) therefore reduces to 

\begin{equation}
	B_{\rm r} = B_{10}f_1(r)\cos \theta.
	\label{eqApp:DipBr}
\end{equation}
where we have chosen to use the $l=1$, $m=0$ mode and note that the Legendre polynomial $P_{10}$ is given by $\cos \theta$. An identical result is obtained for the $l=1$, $m=1$ mode or any combination of the $l=1$ modes but we will proceed with the $l=1$, $m=0$ mode for convenience. In equation (\ref{eqApp:DipBr}), $f_1(r)$ is given by

\begin{equation}
	f_1(r) = \frac{2\tilde{r}^{-3} + \tilde{r}_{\rm ss}^{-3}}{\tilde{r}_{\rm ss}^{-3} + 2}.
	\label{eqApp:ReducedF}
\end{equation}
The flux at a given radial distance form the stellar surface for a pure dipole mode is therefore given by

\begin{equation}
	\Phi_{10}(r) = \oiint\limits_{S} | B_{\rm r}(r) | {\rm d}S	= B_{10} f_1(r) r^2 \int |\cos \theta| \sin \theta {\rm d}\theta \int {\rm d}\phi	= 2\pi B_{10} f_1(r) r^2
\label{eqApp:Flux}
\end{equation}
where $S$ is a spherical surface of radius $r$. Finally, the ratio of the open flux to the surface flux for a pure dipole mode is given by

\begin{equation}
	\frac{\Phi_{10}(r_{\rm ss})}{\Phi_{10}(r_{\star})} = \frac{f_1(r_{\rm ss})}{f_1(r_{\star})} \left(\frac{r_{\rm ss}}{r_{\star}}\right)^2.
	\label{eqApp:FluxRatio}
\end{equation}
Substituting equation (\ref{eqApp:ReducedF}) in, one obtains

\begin{equation}
	\frac{\Phi_{10}(r_{\rm ss})}{\Phi_{10}(r_{\star})} = \frac{\Phi_{\rm open,dip}}{\Phi_{\rm \star,dip}} = \frac{3\tilde{r}_{\rm ss}^2}{2\tilde{r}_{\rm ss}^3 +1}.
	\label{eqApp:FluxRatioFinal}
\end{equation}

\end{document}